\documentstyle[prd,aps]{revtex}
\begin{document}
\input psfig
\pssilent
\title{Consistent discretizations for classical and quantum 
general relativity}
\author{Rodolfo Gambini$^1$, and Jorge Pullin$^2$}
\address{1. Instituto de F\'{\i}sica, Facultad de Ciencias, 
Universidad
de la Rep\'ublica, Igu\'a 4225, CP 11400 Montevideo, Uruguay}
\address{2. Department of Physics and Astronomy, Louisiana State
University, 202 Nicholson Hall, Baton Rouge, LA 70803-4001}
\date{August 23th 2001}
\maketitle
\begin{abstract}
We consider discretizations of the Einstein action of general
relativity such that the resulting discrete equations of motion form a
consistent constrained system. Upon ``spin foam'' quantization of the
system, consistency allows a natural way of recovering the correct
semi-classical theory. A consistent set of approximations to the
Einstein equations could also have implications for numerical
relativity and for the construction of approximate classical
observables for the theory.
\end{abstract}

There has been significant growth in interest in computing the path
integral for general relativity \cite{spinfoam} due to the
introduction of mathematical tools \cite{AsLe} that allow to control
the space of quantum states of the gravitational field. The approach
is generically known as ``spin foam'' approach to quantum gravity. It
consists in postulating discrete approximations to the path integral
of general relativity  in terms of a simplicial
decomposition of space-time. There has been steady progress in showing
that, at least for a given discretization, there is a chance that the
discrete path integral can be finitely computed, even in the
Lorentzian case \cite{CrPeRo}. Part of the motivation for this
approach comes from the success in spin foam approaches to quantize
topological field theories \cite{ooguri}. However, there is a
significant difference between topological field theories and general
relativity. Topological field theories have a finite number of degrees
of freedom, whereas general relativity is a genuine field theory with
local degrees of freedom. This has important implications at the time
of considering the spin foam approach. In topological field theories,
one only needs to consider a discretization on a given simplicial
decomposition, since the results are 
independent of the choice of simplicial decomposition of the
manifold. In general relativity this is not likely to be the case. One
therefore is left with an approximation to the theory that is
discretization dependent. The hope is that in the limit in which one
infinitely refines the triangulation one will recover the continuum
theory through some kind of averaging or limiting procedure. This is
problematic. Since general relativity is not perturbatively
renormalizable, there is no guiding principle that ensures that the
limit will be well defined. For instance, in Yang-Mills theories on
the lattice, one can write infinitely many discretizations of the
dynamics that are all different but become the same due to the phase
transition of the theory that defines the continuum limit. No such
procedure is known for gravity. In fact, a point of view advocated
some times in the gravitational case is that one should not take the
continuum limit since the theory has a natural cutoff length given by
the Planck length. This point of view suggests that quantum gravity
will be inherently a discrete theory (see for instance the discussion
in \cite{CrPeRo}). Similar comments apply to the canonical quantization
of general relativity. In this case, proposals for the Hamiltonian
constraint have been made \cite{Qsd}, that are based upon
discretizations of the theory. Different discretizations appear to
lead to different theories \cite{GaRo}, which may or may not yield the
same semiclassical limits, which are currently beginning to be studied
\cite{coherent}

The above comments suggest that either due to fundamental reasons or
due to practical ones, one will be dealing with discretizations of
general relativity at the time of quantizing the
theory. Discretizations are also useful in a completely different area
of physics: numerical simulations of the Einstein equations. Again,
one is really interested in the limit in which the discretization is
infinitely refined, but in practice one handles a discrete theory when
performing a numerical simulation.

A striking element of most discretizations of general relativity
currently used is that in and of themselves, they define {\em
inconsistent} theories. For instance, it would be highly desirable if
the spin foam models of the path integral stemmed from the
quantization of discretizations of the classical action. However, it
is quite non-trivial to obtain discretizations of the action of
general relativity such that the classical equations of motion
stemming from the discrete action have non-trivial solutions and
reproduce in the limit the continuum theory at the level of the
equations of motion (for a discussion in the context of Regge Calculus
see \cite{MiBR}). In the case of numerical relativity, the set of
discrete equations that represent the Einstein evolution equations are
usually inconsistent with the discrete equations that represent the
constraints. It is a bit surprising that these significant
difficulties have not attracted attention in the past, but the general
attitude towards discretizations has been that whatever
inconsistencies one has will disappear in the continuum limit and are
therefore tolerable. However, if the continuum limit is not achieved
(either by fundamental or practical reasons as we discussed above), we
believe these inconsistencies should be taken seriously.

The inconsistencies might appear surprising at first, but they are
not. There is a body of literature dealing with the space-time
discretization of dynamical systems \cite{dismech}, and even for
elementary mechanical systems it is difficult to achieve
consistency. As an example, let us consider a simple mechanical
system, a particle in a potential, and let us take time to be
discrete. The action then reads $S=\sum_{i=1}^{N} p_i (q_{i+1}-q_i) +
\left[{p_i}^2/2 +V(q_i)\right]\Delta t$. It one works out the discrete
equations, one gets $q_{i+1} -q_i = p_i$,
$p_{i}-p_{i-1}=-V'(q_i)$. Notice the asymmetry in the left
members. This time asymmetry, in turn, implies that energy is not
conserved, as can be checked through a straightforward
calculation. This problem also implies that if one adds constraints
that are not holonomic (i.e. they mix $p_i$ and $q_i$) the constraints
are not preserved in evolution. Most of the literature up to now on
``discrete mechanics'' has considered only systems with holonomic
constraints (see \cite{WeMa} for references). Such systems therefore
do not include general relativity.

In this paper we would like to suggest two new proposals for
constructing consistent discretizations of general relativity. The
first one will be in the context of discretizing equations of motion
in a consistent way, like one would do in numerical relativity. Here
what we will seek to find is discrete systems of evolution equations
that are compatible with the discrete constraints. The
second will be in discretizing an action like is done in spin foam
quantum gravity. Here what we will attempt to do is to find a
discretization of the classical action that yields discrete equations
of motion that are a good approximation to the continuum equations of
motion and that admit solutions that approximate the classical
solutions. We will argue how this will help in constructing the
semi-classical limit of the theory.

{\em Discretizing the Einstein equations}. Let us start by considering
how to discretize consistently the equations of general relativity. To
motivate what we will propose, we would like first to work out a
simpler example, closely related to one solved by T.D. Lee
\cite{TDLee}. We will see that the technique introduced naturally
suggests what to do in the case of general relativity.  The example is
a one dimensional mechanical system.  The discrete action for the
system is $S = \sum_n p_n (q_{n+1}-q_n) -H(p_n,q_n)(t_{n+1}-t_n)$.
The novelty in T.D. Lee's approach is to consider the $t_n$ as
canonical variables, the theory being ``discretely parameterized'' by
$n$.  Variation with respect to $t_n$ shows that $H_{n}=H_{n-1}$ and
therefore the Hamiltonian is automatically conserved. One has three
equations of motion, given by variations with respect to
$t_n,q_n,p_n$. The equations will determine $t_n,p_n,q_n$ and
therefore the time intervals are determined by the evolution.  Let us
now consider the same construction for a constrained system. Consider
the same system with a constant energy $E_0$. The action will read $S
= \sum_n p_n (q_{n+1}-q_n) -\lambda_n (H(p_n,q_n)-E_0)$. We have
incorporated into the Lagrange multiplier $\lambda_n$ the time
interval. We now have an equation of motion with respect to
$\lambda_n$ which states that the energy is constant and equations of
motion with respect to $q_n,p_n$ that will involve the Lagrange
multiplier. Solving the system shows that the Lagrange multiplier will
be determined by the evolution. Is time reparametrization lost? It is
not, since we had absorbed the time interval into the definition of
the multiplier, we can always choose a different time interval in the
theory. This example suggests what we can do in general
relativity. Our proposal consists in taking the $3+1$ decomposition of
the Einstein equations discretized on a grid, like it is normally done
in numerical relativity. One will get discrete time evolution
equations for the extrinsic curvature and the three metric. The right
hand sides of these equations will involve the (discretized) lapse and
shift. We then add to these equations the four discrete constraint
equations evaluated at time $n+1$. The resulting system should be
solved for the metric, extrinsic curvature, lapse and shift, in such a
way that both evolution and constraint equations are solved. One is
then left with a consistent system, at the expense of fixing the lapse
and shift. It might appear paradoxical at first that the lapse and
shift are fixed, but it is not. Since discrete theories do not have
diffeomorphism invariance, in general one has to fix the lapse and
shift for the theories to be consistent. This was observed in the
context of Regge Calculus by Friedmann and Jack \cite{FrJa}. However,
it still seems strange that one is trying to approximate a continuum
theory, which has arbitrary lapse and shifts, with a discrete theory
that has fixed ones. The answer lies in the fact that when one
discretizes a theory there is implicit a mapping between space-time
and the discrete lattice. Such mapping is arbitrary and restores the
reparametrization freedom of general relativity. It is analogous to
what happened in the case of the harmonic oscillator, where what was
really fixed was not the Lagrange multiplier, but its value times the
discretized time interval. It is evident that what we have up to now
is just a proposal, and further exploration in practical numerical
implementation will be needed. In particular, although the system
constructed solves exactly the discrete constraints (and therefore
approximates well the continuum constraints forever in time) our
experience with model systems shows that since one needs to solve
non-linear equations for the Lagrange multipliers, there are
situations where one might only find complex solutions, which will
render the scheme useless. Studies applied to one dimensional systems
are currently under way. After completion of this manuscript we became
aware of work of Van Putten along similar lines \cite{MVP}, who 
appears to have tested a closely related proposal in the context of
Gowdy waves.

{\em Discretizing the action of general relativity and quantization}

In the spin foam approach to quantum gravity, a certain discretization
of the path integral is proposed. This is not necessarily achieved by
a unique prescription. In general what people have in mind would be to
write the action of general relativity using connection-type
variables, discretized on a given simplicial decomposition. One then
performs the functional integrals on the connections and is left with
an expression that is given by a sum over ``spin foams'', that is, an
assignment of group-valued labels to a graph. Due to the
discretization, the functional integrals become ordinary integrals on
the group (typically $SU(2)$). This ``procedure'' is loosely
followed. Usually, certain expressions for the final discretization
are chosen a priori based on certain desirable criteria, they are not
necessarily obtained strictly through a discretization of the
classical action. As we mentioned in the introduction, the resulting
quantum theories are discretization-dependent and it is not clear in
what sense they are connected with classical general relativity. It is
expected that in some limiting procedure one should recover the
semi-classical theory. There is however no guiding principle to take
this limit.

Our point of view is that it would be highly desirable to construct
spin foam models starting with a consistent discretization of the
action of general relativity, in the sense that before quantization,
one has a discrete classical action that yields discrete equations of
motion that are consistent and with solutions that in the continuum
limit go to those of general relativity. Even with this scenario, one
should be aware that discretized theories in general have more
solutions that the continuum theory. These solutions have ``spikes''
or ``high frequency'' oscillations and do not converge when one
refines the discretization to solutions of the continuum theory. 

The strategy for computing the functional integral will then be to
choose a consistent discretization and compute the transition
amplitude for it; this calculation is well defined since one is
dealing with a system with a finite number of degrees of freedom and a
discrete time. Therefore the transition amplitude is given by an
ordinary integral. If we were to stop here and try to use the
transition amplitude as an approximation to the path integral of the
continuum theory, we would quickly discover that one has a bad
approximation. Since the discrete theory contains a large number of
spurious solutions, the resulting transition amplitude will in general
fail to capture the behavior one associates with the continuum
theory. Moreover, the result will be discretization-dependent. This is
the typical situation that is being confronted by current approaches
to spin-foam quantization. In our approach we have a clear
advantage. Since the discretization constructed is consistent, one is
ensured that the discrete theory will contain a solution that
approximates better and better the continuum solution as one refines
the discretization. It does, however, contain many other spurious
solutions. To get rid of these we propose the following
construction. Consider the discrete theory on a simplicial
decomposition of the manifold with a finite number of simplices. We
then average the resulting transition amplitude over all possible
discretizations with up to the given number of simplices. This is a
finite sum, so it is well defined. Each discretization will therefore
contain a transition amplitude that approximates the one of the
continuum theory and many others that are discretization
dependent. The averaging procedure will therefore suppress the
discretization-dependent amplitudes, since one in general has only one
of each per discretization, and will be dominated by the amplitudes
that approximate those of the continuum theory. One can then repeat
this averaging with more simplices. As one increases the number of
simplices, given that the discretization is consistent, one ensures
that the transition amplitude will model better and better that of the
continuum theory. What would have happened if one had picked an
inconsistent discretization? In such case, the various approximations
correspond to theories that are not related to the classical theory in
any obvious way. Although one can perform a path integral quantization
using an action that yields inconsistent equations of motion, we
cannot see a sense in which summing a large number of such
contributions will conspire to yield an amplitude related to classical
general relativity, unless one could find a guiding principle that
could suggest this should happen (e.g. renormalizability).

Let us discuss a consistent discretization suitable for spin foam
quantization. The discretization has been proposed by Reisenberger
\cite{Re}. The new elements in this paper is to indicate that indeed
the discretization is consistent at least for simple
discretizations. Although the discretization of Reisenberger was
introduced for Euclidean gravity, it can be generalized to the
Lorentzian case. The functional integrals in that case are problematic
since the group is non-compact, but the recent results of
\cite{CrPeRo} suggest that they could be handled through a certain
regularization. The model is based on the Plebanski \cite{Pl} action
of general relativity,
\begin{equation}
S=\int \Sigma_i\wedge F^i - {1 \over 2} \phi^{ij}\Sigma_i \wedge
\Sigma_j
\end{equation}
where $\phi^{ij}$ is a symmetric traceless matrix, $F^i$ and
$\Sigma_i$ are $SU(2)$ valued two forms. $F^i$ is the curvature of an
$SU(2)$ connection. Variation with respect to $\phi^{ij}$ gives a
constraint equation for the $\Sigma_i$ that implies (at least for
non-degenerate solutions that it is given by a skew product of
one-forms $\Sigma^i = e_j \wedge e_k \epsilon^{ijk}$. If one
substitutes back in the action, one recognizes the Samuel \cite{Sa}
action for general relativity. One now considers a simplicial
decomposition of the manifold $\Delta$. Each 4-simplex $\nu$ is
bounded by five tetrahedra $\tau$, intersecting at ten faces $\sigma$.
One can then introduce the ``derived'' (or ``dual'') complex, formed
by wedges determined by the following four points: the simplex center
$C_\nu$, the centers of two adjacent tetrahedra $C_\tau$, and the
centers of the faces $C_\sigma$. Each wedge $s$ is bounded by a
polygonal line starting at the center of the four simplex, going over
to the center of one of the tetrahedra, continuing to the center of a
face, over to the center of the other tetrahedra adjoined by that face
and returns to the center of the 4-simplex. There is therefore one
wedge ``dual'' to each face. To introduce the discrete variables, the
connection is represented by the holonomy along one of the sides of
the wedge, the curvature by taking the trace of the product of a
closed holonomy along the wedge times a generator of $SU(2)$, which we
call $\theta^i$. Each wedge has associated with it an object $e_{si}$
that represents the two form $\Sigma_i$. The matrix $\phi^{ij}$ is
assumed based at the center of the 4-simplex, so we denote it
$\phi^{ij}_\nu$ The discrete action is given by \cite{Re},
\begin{equation}
S_\Delta= \sum_{\nu\in\Delta} \left[\sum_{s\in \nu} 
e_{si} \theta^i -
{1 \over 60} \phi^{ij}_\nu \sum_{s,\bar{s}\in\nu} e_{si} e_{\bar{s}j} 
{\rm sgn}(s,\bar{s})\right]
\end{equation}
where ${\rm sgn}(s,\bar{s})$ is non-vanishing if the two wedges share
at most one vertex, and is $\pm 1$ depending on the orientation of the
wedges. 

A wedge has two types of links, the ones that start from the center of
the simplices (``interior links'') and the ones  that go from the
center of the tetrahedra to the center of the face (``boundary
links'').  The discrete equations of motion are obtained by varying
with respect to the holonomies associated with the two types of links,
with respect to $\phi^{ij}_\nu$ and with respect to the
$e_{si}$. Varying with respect to the interior links one gets a sum
over all the faces of the quantity $e_{is}$ associated with the wedge
dual to the face, parallel-transported to the center of the
simplex. This sum has to vanish and corresponds to a discretization of
Gauss' law. Varying with respect to the boundary links one gets a
holonomy around the wedge with the insertion of an $e_is$ with $s$ the
dual of the wedge. The equation states that this quantity is the same
for all wedges associated with simplices that share the face $\sigma$.
Varying with respect to $\phi^{ij}$ one gets a constraint that
corresponds in the continuum to the two form $\Sigma$ being
decomposable into the wedge product of two one-forms. Finally, varying
with respect to $e_{is}$ one gets an expression that links the
``curvature'' $\theta$ to the multipliers $\phi^{ij}$ and the field
$e_{si}$. In the continuum this would correspond to relating the
curvature with the Weyl tensor. As we have suggested, all the
equations can be proved in detail to 
have the correct continuum limits \cite{Re}.

Is the resulting theory consistent? In order to establish this in
general one would need a full canonical analysis, which is not 
available at present. We have however been able to establish that the
equations of motion are consistent for particular discretizations. As
an example let us consider a single simplex. To count variables, we
start by choosing a gauge in which the only non-trivial holonomies are
associated to the boundary links that are outgoing from the face dual
to the wedge on which the link lives. There are 10 such holonomies,
and in terms of them we can compute all holonomies along wedges. 
To see that this gauge exists, one first chooses a gauge
transformation at the centers of tetrahedra that makes the internal
holonomies identity. One then chooses a gauge transformation in the
center of faces in such a way that the incoming holonomies are trivial
(one could alternatively choose the outgoing ones). We therefore have
30 variables associated with the holonomies (for $SU(2)$ each 
holonomy can be parameterized by three real variables). We also have
the five $\phi^{ij}$'s and 30 $e_{si}$'s, for a total of 65
variables. Counting equations, the Gauss law gives 15 identities
(three per each of the five faces). The second set of equations is
trivial, since there is only one simplex. The third set of equations
gives five conditions and the last set corresponds to 30 conditions. 
So we have a total of 50 equations for the 65 variables. The rest
corresponds to ``boundary data'' (the boundary of the tetrahedra can
be viewed as a sandwich-boundary value problem) that are freely
specifiable. Similar countings can be performed for more complex
simplicial decompositions. We have failed to find one that leads to
inconsistencies.

We therefore see that we have two different discretizations of
classical general relativity that are consistent, one more geared
towards a numerical implementation and one suited for spin foam
quantization. We believe that the use of consistent discretizations
will prove fruitful in producing theories with the correct continuum
limits, both at a classical and quantum level. Having a consistent
discrete theory allows to perform studies of a concrete, definitive
nature (for instance computing observables), providing a laboratory
for testing the validity of the approximations constructed.

We wish to thank Doug Arnold and Richard Price for comments. 
This work was supported by grants NSF-PHY090091, NSF-INT-9811610.

\end{document}